\documentclass[aps,preprint,superscriptaddress]{revtex4}
\usepackage{color}
\usepackage{amssymb}
\usepackage{color}
\usepackage{longtable}
\usepackage{graphicx}
\usepackage{amsmath}
\usepackage{amsfonts}
\usepackage{amssymb}
\usepackage{natbib} 
\usepackage{setspace}
\usepackage{xspace}
\usepackage{cancel}
\usepackage{physics}
\usepackage{tensor,braket,color,bbold,dsfont,slashed}
\usepackage{newtxmath}
\usepackage{mathrsfs}
\usepackage[normalem]{ulem}
\usepackage[colorlinks=true, pdfstartview=FitV, bookmarks=true, bookmarksnumbered=true, breaklinks]{hyperref}
\usepackage{mathtools,braket}
\usepackage{soul}
\usepackage{lipsum}  
\usepackage{color}
\definecolor{blue}{rgb}{0.0, 0.0, 1.0}
\definecolor{red}{rgb}{1.0, 0.0, 0.0}
\definecolor{royalblue}{rgb}{0.0, 0.14, 0.4}
\definecolor{myblue}{rgb}{0.0, 0.4, 0.6}
\hypersetup{linkcolor=royalblue, citecolor=blue, urlcolor=royalblue}

\usepackage{hyperref}
\hypersetup{colorlinks=true,citecolor=blue,linkcolor=blue,urlcolor=blue}

\usepackage[mathlines]{lineno}
\usepackage{tikz,xcolor,hyperref}
\definecolor{lime}{HTML}{A6CE39}
\DeclareRobustCommand{\orcidicon}{%
	\begin{tikzpicture}
	\draw[lime, fill=lime] (0,0) 
	circle [radius=0.16] 
	node[white] {{\fontfamily{qag}\selectfont \tiny ID}};
	\draw[white, fill=white] (-0.0625,0.095) 
	circle [radius=0.007];
	\end{tikzpicture}
	\hspace{-2mm}
}

\foreach \x in {A, ..., Z}{%
	\expandafter\xdef\csname orcid\x\endcsname{\noexpand\href{https://orcid.org/\csname orcidauthor\x\endcsname}{\noexpand\orcidicon}}}

\begin{document}
\title{Effects of Symmetry Energy on the Equation of State\\for Hybrid Neutron Stars}

\author{Parada~T.~P.~Hutauruk}
\affiliation{Department of Physics, Pukyong National University (PKNU), Busan 48513, Korea}
\affiliation{Department of Physics Education, Daegu University, Gyeongsan 38453, Korea}

\author{Hana Gil}
\affiliation{Department of Physics Education, Daegu University, Gyeongsan 38453, Korea}
\affiliation{Center for Extreme Nuclear Matters, Korea University, Seoul 02841, Korea}

\author{Seung-il Nam}
\affiliation{Department of Physics, Pukyong National University (PKNU), Busan 48513, Korea}
\affiliation{Center for Extreme Nuclear Matters, Korea University, Seoul 02841, Korea}
\affiliation{Asia Pacific Center for Theoretical Physics (APCTP), Pohang 37673, Korea}

\author{Chang Ho Hyun}
\affiliation{Department of Physics Education, Daegu University, Gyeongsan 38453, Korea}
\affiliation{Center for Extreme Nuclear Matters, Korea University, Seoul 02841, Korea}
\date{\today}

\begin{abstract}
In this paper, the implications of the symmetry energy on the hadron and quark phase transitions in the compact star, including the properties of the possible configurations of the quark-hadron hybrid stars, are investigated in the frameworks of the energy-density functional (EDF) models and the flavor SU(2) Nambu--Jona-Lasinio (NJL) model with the help of the Schwinger's covariant proper-time regularization (PTR) scheme. In this  {theoretical setup}, the equations of states (EoSs) of hadronic matter for various values of symmetry energies obtained from the EDF models are employed to describe the hadronic matter, and the  {flavor} SU(2) NJL model with various repulsive-vector interaction strengths are used to describe the quark matter. We then observe the obtained EoS in the mass-radius properties of the hybrid star configurations for various vector interactions and nuclear symmetry energies by solving the Tolman-Oppenheimer-Volkoff equation. We obtain that the critical density at which the phase transition occurs varies over the density (3.6--6.7)$\rho_0$ depending on the symmetry energy and the strength of the vector coupling $G_v$. The maximum mass of the neutron star (NS) is susceptible to $G_v$. When there is no repulsive force, the NS maximum mass is only about $1.5M_\odot$,
but it becomes larger than $2.0M_\odot$ when the vector coupling constant is about half of the  {attractive} scalar coupling constant. Surprisingly, the presence of the quark matter does not affect the canonical mass of NS ($1.4M_\odot$), so observing the canonical mass of NSs can provide unique constraints to the EoS of hadronic matter at high densities.
\end{abstract}

\maketitle

\section{Introduction} \label{sec:intro}

A key question about the composition of matter at densities several times the nuclear saturation density ($\rho_0$) is how it differs from the state that constitutes the atomic nuclei. It was widely known that several exotic states are candidates for the states of matter at ultra-high densities: The creation of hyperons, the onset of Bose-Einstein condensates, and transformation to quark matter. Using theories of subatomic physics, one can estimate in what conditions the new state of matter will emerge. However, the existence of the new state matter is not yet established, for instance, the appearance of the hyperons and $\Delta$ baryon in the matter and how they interact with other matter constituents are still open questions.

In a recent work of Ref.~\cite{epja2022}, the in-medium interaction of the $\Lambda$ hyperon was investigated within non-relativistic nuclear density functional theory. Parameters for the interaction of $\Lambda$ hyperon were determined to reproduce the single-$\Lambda$ hypernuclear data,  and the effect of the density dependence of the symmetry energy was examined in more systematic detail. It was shown that the density at which $\Lambda$ hyperons  {appear} depends strongly on the symmetry energy. However, the appearance density of hyperons is always larger than $3\rho_0$, so the hyperon is not likely to affect the properties of a canonical mass ($1.4M_\odot$) of NS because the density at the center of the NS canonical mass is generally less than $3\rho_0$. Kaon condensation was also considered in the relativistic and non-relativistic models~\cite{kcond1, kcond2}. Unless the interaction of the kaon in dense nuclear  {matter} is super-strongly attractive, the formation of the kaon condensation is less probable.

Analogously, the matter composition in NSs naively can be illustrated as the quarks in the baryons on a microscopic scale. In the bag model, baryons are described as spherical bags in which quarks are assumed to be confined. In the bag model, the charge radius of the proton is expected around 0.6 -- 0.8 fm (corresponding to the volume 0.9 -- 2.1 fm$^3$). The inverse of the number density ($\rho$) is the volume occupied by a nucleon.  If $\rho_0 = 0.16\, {\rm fm}^{-3}$, one nucleon occupies 6.25 fm$^3$ at the saturation density. This volume is larger than the volume of a nucleon, so nucleons are separated spatially. At densities in the range of 3 -- 6 $\rho_0$, the volume per nucleon becomes equal to the volume of a nucleon, so the nucleons begin to overlap each other.
As the matter density evolves above the onset of overlapping,  {nucleons} overlap more, and it becomes vague to define the confinement of a quark in a specific bag. Matter transforms to the phase of deconfined quarks. A similar condition is expected to occur in NS.

Given the equation of state (EoS) of dense matter, one can calculate the bulk properties of the {NSs} such as mass, radius, and particle distribution in the core by solving the Tolman-Oppenheimer-Volkoff (TOV) equations. With the nuclear models that are well constrained by the nuclear properties and nuclear matter properties from the heavy ion collision experiments, {\it ab initio} calculations, and modern neutron star observation data~\cite{Papakonstantinou:2016zpe,kids-nuclei1,kids-nuclei2,kids-sym,kids-k0,kids-ksym}, the density at the center of a canonical mass of NS is obtained to be about or less than $3\rho_0$, and that of the heaviest stars ($\geq 2 M_\odot$) is around $6\rho_0$. The degrees of freedom that constitute the core of NSs might likely change from hadron to quarks
in  {NSs} with masses in the range ($1.4 - 2.0$)$ M_\odot$.

The main issue of the present work is to explore the uncertainty of the critical density ($\rho_c$) at which the transition from the hadronic phase to the deconfined quark matter begins and its impact on the properties of the NS. For this purpose, for the EoS of hadronic matter ({HM}), we employ the KIDS (Korea-IBS-Daegu-SKKU) energy density functional (EDF) in which the density dependence of the symmetry energy is calibrated to the radius of $1.4M_\odot$ mass NSs in the range $R_{1.4} = 11.8 -12.5$ km \cite{kids-k0}.  {The KIDS0, KIDS-A, KIDS-B, KIDS-C, and KIDS-D models}, as well as the standard Skyrme SLy4 model used in the work, have distinctive density dependence of the symmetry energy, so the results will provide the range of $\rho_c$ within the uncertainty of the nuclear matter EoS. On the quark matter (QM) side, the proper time NJL model is used for the EoS of the deconfined quark state. As discussed above, massive NSs are likely to have a QM phase in the core, so we investigate the range of parameters in the NJL model that are compatible with the NS mass larger than $2M_\odot$. As a result, we can determine the ranges of the parameters in the QM EoS. Combining the EoSs of the KIDS model with those of the NJL model, we determine the range of $\rho_c$
from the condition $P_{\rm HM}(\rho_c) = P_{\rm QM}(\rho_c)$ at $\rho_{\rm HM} = \rho_{\rm QM} = \rho_c$, where $P_{\rm HM}$, $P_{\rm QM}$, $\rho_{\rm HM}$, and $\rho_{\rm QM}$ represent the pressure in the hadronic phase, pressure in quark phase, density in hadronic phase, and density in quark phase, respectively. The results will shed light on 
(i) the effect of the symmetry energy on the change of phase in the  {NS} core,
(ii) consistency of the hybrid models with the NS properties determined from astronomical observations, and
(iii) uncertainty of the EoS in the hadronic and quark phases.

This work is organized as follows. In Sec.~\ref{sec:hqmeos}, models for the hadronic matter and deconfined quark matter are briefly introduced. The hadronic matter is described in the KIDS-EDF model, whereas the quark matter is described in the  {flavor} SU(2) Nambu--Jona-Lasinio model with the help of the Schwinger proper-time regularization scheme, that is so-called NJLPTR model. We present and discuss the numerical result of the current work in Sec.~\ref{sec:NR}, and we summarize the work in Sec.~\ref{sec:SUM}.

\section{Formalism for the hadron and quark phases} \label{sec:hqmeos}

\subsection{Hadronic matter from the KIDS functional} \label{sec:hm}

In this section, we first present the description of  {HM} in the KIDS-EDF model~\cite{Papakonstantinou:2016zpe,Hutauruk:2022bii,Hutauruk:2022bso}. This model was constructed based on the Fermi momentum expansion, which is relevant for infinite nuclear matter (INM) and an NS. The energy per particle for homogeneous hadronic matter in the powers of the cubic root of the density is 
given by
\begin{eqnarray}
    \label{kids1}
        \mathcal{E}_{\rm  {HM}} (\rho,\delta) &=& \mathcal{T} (\rho,\delta) + \sum_{j=0}^{3} (\alpha_j + \beta_j \delta^2) \rho^{(1+j/3)},
\end{eqnarray}     
where $\mathcal{T} (\rho,\delta)$ is the kinetic energy. $\delta = (\rho_n -\rho_p)/\rho$ and $\rho = \rho_n +\rho_p$ are respectively the isospin asymmetry and baryon density. The $\alpha_j$ and $\beta_j$ are respectively the coefficients for the symmetric and antisymmetric nuclear matter.

The symmetry energy $S(\rho)$ is straightforwardly determined through the second derivative of the energy  {per particle} with respect to $\delta$ at  {the} nuclear saturation density. It then gives
\begin{eqnarray}
    \label{kids2}
    \mathcal{E}_{\rm  {HM}} (\rho,\delta) &=& \mathcal{E} (\rho,0) + S(\rho) \delta^2 + \mathcal{O} (\delta^4),\\
    S (\rho) &=& \frac{\hbar}{6M} \left(\frac{3\pi^2}{2}\right)^{2/3} \rho^{2/3} + \sum_{j=0}^3 \beta_j \rho^{(1+j/3)},
\end{eqnarray} 
where $M\equiv(M_n + M_p)/2$ is the nucleon mass in free space. The pressure for the  {HM} can be calculated from the energy per particle with respect to baryon density. It then 
has the form
\begin{eqnarray}
    \label{kids3}
        P_{\rm  {HM}} &=& \rho^2 \frac{\partial \mathcal{E}_{\rm  {HM}}(\rho,\delta)}{\partial \rho}.
\end{eqnarray}  
  {Here we emphasize again that}, in this work, the HM is represented by the five KIDS-EDF models: KIDS0, KIDS-A, KIDS-B, KIDS-C, and KIDS-D models, and  {the} Skyrme force SLy4 model. Those various EDF models have different nuclear symmetry energies, which cover soft to stiff nuclear symmetry energies. 
\begin{figure}
    \centering
    \includegraphics[width=0.95\textwidth]{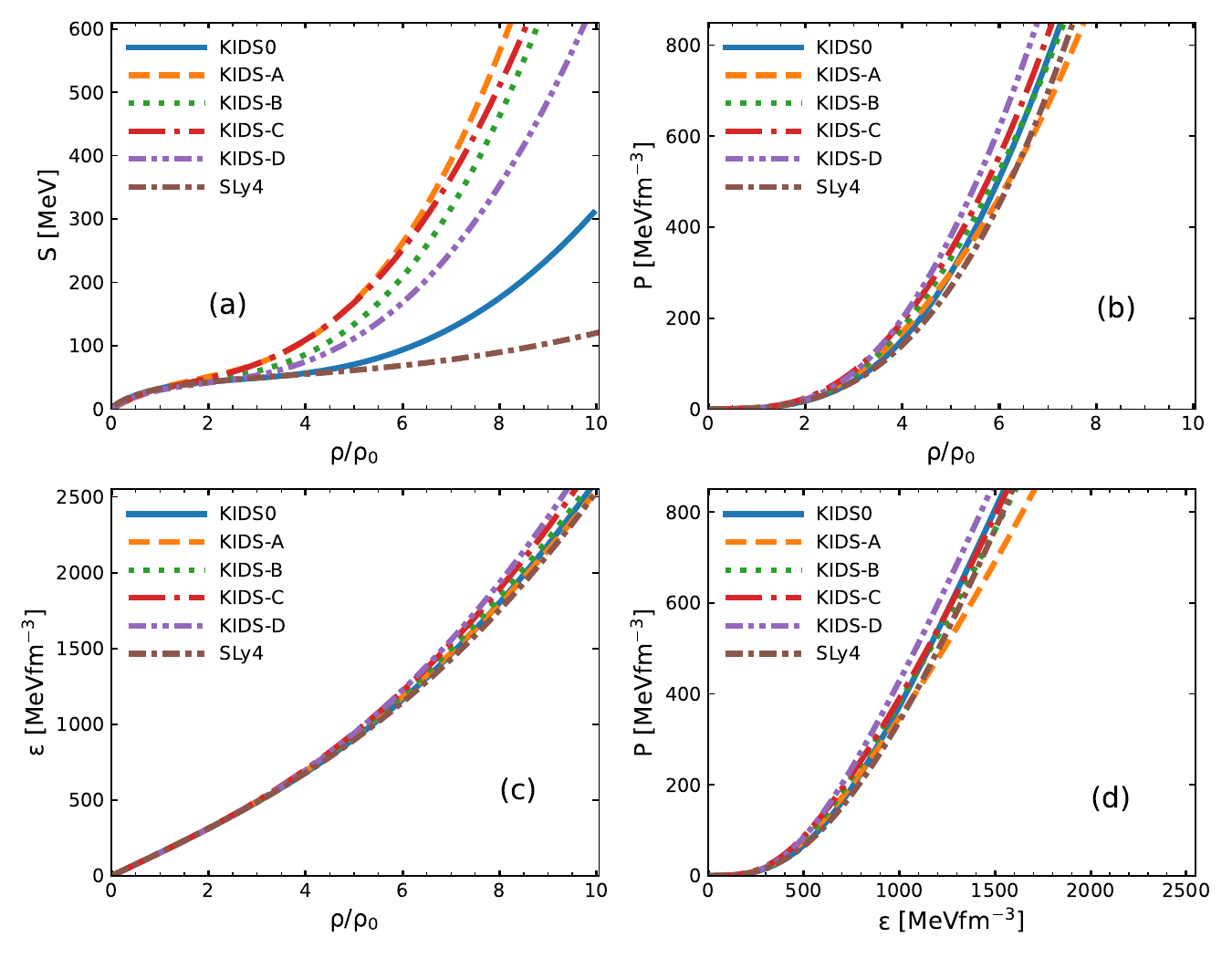}
    \caption{ {From the KIDS EDFs and the SLy4 model: (a) Symmetry energy, (b) Pressure of HM as a function of $\rho/\rho_0$, (c) Energy density of HM as a function of $\rho/\rho_0$ and (d) EoS of HM}.}
    \label{fig1}
\end{figure}
The nuclear symmetry energy results for the KIDS0, KIDS-A, KIDS-B, KIDS-C, KIDS-D, and SLy4 models are given in Fig.~\ref{fig1} (a). It clearly shows different ranges of symmetry energies covering soft and stiff symmetry energies that will be used to construct the EoS of the hybrid model.  Figure~\ref{fig1} (a) shows, for HM, the stiffest symmetry energy is given by the KIDS-A model, and the softest is given by the SLy4 model. We also provide the pressure of the  {NS} matter as a function of the densities in Fig.~\ref{fig1} (b).  The pressures of the symmetric matter obtained from the EDF models are consistent with the heavy-ion collision (HIC) reaction data at high density~\cite{Danielewicz:2002pu} as well as the chiral perturbation theory (ChPT) at low density~\cite{Tews:2012fj}.

The result for the energy densities $E_{\rm HM}$ of the NS matter for the hadronic models with respect to density is given in Fig.~\ref{fig1} (c). Results for the $P_{\mathrm{HM}}$-$E_{\mathrm{HM}}$ relation for the hadronic models are shown in Fig.~\ref{fig1} (d). This EoS relation is required as input for calculating the TOV equation to obtain the properties of the NSs. In the HM model, we found that the $M_\mathrm{NS}/M_{\odot}$ results for the KIDS0 and SLy4 models are rather similar because both models have soft symmetry energies. In contrast, the KIDS-A and KIDS-B models have stiff symmetry energies giving a larger radius of the NS. Overall, the mass and radius of NS results for all KIDS models fit well with the recent observations~\cite{Antoniadis:2013pzd,NANOGrav:2019jur,Demorest:2010bx}.

\subsection{Pure quark matter from the NJLPTR model} \label{QM}

In this section, we describe the pure (nonstrange) quark matter (PQM) in the framework of the flavor SU(2) Nambu--Jona-Lasino (NJL) model~\cite{Hutauruk:2021kej,Tanimoto:2019tsl,Hutauruk:2021dgv,Bentz:2001vc,Baym:2017whm,Buballa:2003qv,Glendenning:1992vb,Hutauruk:2022sbm,Hutauruk:2016sug,Hutauruk:2018zfk}. This model is built in terms of the quark degrees of freedom in the form of the four-fermion contact interactions. Therefore, this model is very suitable and powerful for the description of quark matters. The general NJL Lagrangian density for two quark flavors can be written as
\begin{eqnarray}
    \label{qmnjl1}
    \mathscr{L}_{\mathrm{NJL}} &=& \bar{\psi}_q \left(i \partial\!\!\!/ - \hat{m}_q \right)\psi_q + G_s \left[(\bar{\psi}_q \psi_q)^2 +(\bar{\psi}_q\gamma_5 \vec{\tau} \psi_q)^2 \right]
    - G_\omega (\bar{\psi}_q\gamma_\mu \psi_q)^2 - G_\rho (\bar{\psi}_q \gamma_\mu \vec{\tau} \psi_q)^2, \nonumber \\
\end{eqnarray}
where the quark field is given by $\psi_q = (\psi_u,\psi_d)^T$, $\vec{\tau}$ stands for the Pauli isospin matrices, and $\hat{m}_q = \textrm{diag}[m_u, m_d]$ is the current (bare) quark mass, where, in this work, we use the isospin symmetry, giving $m_u \simeq m_d$. The $G_s$, $G_\omega$, and $G_\rho$ are respectively scalar, isoscalar-vector, and isovector-vector coupling constants. Note that the NJL model suffers the divergence in the quark propagator, therefore, we must apply  { a regularization scheme} to cure the divergence problem. In this work, we perform the  {PTR} scheme to tame the divergence~\cite{Schwinger:1951nm}. In the standard NJL formalism, the  {effective quark mass} can be easily obtained in the mean-field approximation (MFA), which is explicitly given in the  {PTR} scheme as
\begin{eqnarray}
    \label{qmnjl2}
    M_q &=& m_q - 2 G_s \langle \bar{\psi}_q \psi_q \rangle = m_q + \frac{3G_s M_q}{\pi^2} \int_{\Lambda_{\mathrm{UV}}^{-2}}^{\infty} d\tau \frac{\exp[-\tau M_q^2]}{\tau^2},
\end{eqnarray} 
where $\langle \bar{\psi}_q \psi_q\rangle$ is the chiral condensate for flavor $q$, which relates to the order parameter of the spontaneous breaking of chiral symmetry (SB$\chi$S). It is worth noting that the  {effective quark mass} in Eq.~(\ref{qmnjl2}) has an additional correction in the quark matter, which comes from the density term in the quark propagator.

In the  {QM}, the NJL Lagrangian density in Eq.~(\ref{qmnjl1}) is modified by adding an extra term of quark density or quark  {operator} chemical potential. It then has the form
\begin{eqnarray}
    \label{qnjl3}
    \mathscr{L}_{\mathrm{NJL}} &\rightarrow& \mathscr{L}_{\mathrm{NJL}} + \bar{\psi}_q \hat{\mu}_q \gamma^0 \psi_q,
\end{eqnarray} 
where $\hat{\mu}_q = \textrm{diag}[\mu_u, \mu_d]$ is the quark chemical potential matrix. Thus, considering the density effect, the quark propagator for flavor $q$ in momentum space can be written by
\begin{eqnarray}
    \label{qnjl4}
    S_q (p) &=& \frac{1}{\left[\left(p_0 + \tilde{\mu}_q \right) \gamma^0 - \vec{p} \cdot \vec{\gamma} - M_q \right]},
\end{eqnarray}
where the $\tilde{\mu}_q$ is the so-called the reduced quark chemical potential for flavor $q$ and it can can be written as
\begin{eqnarray}
    \label{qnjl5}
    \tilde{\mu}_q &=& \mu_q - 2 G_v \rho_q^{v},
\end{eqnarray}  
where $\rho_q^{v} \simeq \langle \psi_q^{\dagger} \psi_q \rangle = p_{F_{i}}^3/\pi^2$ stands for the individual quark number densities of flavor $q$,  {where $p_{F_{i}}$ is the Fermi momentum for quark flavor $i=(u,d)$}. For simplicity, in the present work, we opt  {the vector coupling} $G_v=G_\omega=G_\rho$ by considering vector dominance and small mass differences between $\rho$ and $\omega$ masses~\cite {Hell:2014xva,Klimt:1990ws}. Also, it is worth noting that the $G_v$ will be treated as a free parameter in this work.

In the standard method, the effective potential for the nonstrange  {QM} in the NJL model is given by~\cite{Lawley:2005ru}
\begin{eqnarray}
    \label{qnjl6}
    \mathscr{V}^{\textrm{QM}}_{\textrm{NJL}}(M_q,\mu_q) &=& 2iN_c \sum_{q=[u,d]}\int \frac{d^4k}{(2\pi)^4} \log \left[ \frac{k^2-M_q^2 + i \epsilon}{k^2-M_{0}^2  -i \epsilon} \right] \nonumber \\
    &+& \sum_{q=[u,d]} \frac{(M_q-m_q)^2}{8G_s} - \sum_{q=[u,d]}\frac{(M_{0}-m_q)^2}{8 G_s} \nonumber \\
           &-& 2 N_c \sum_{q=[u,d]} \int \frac{d^3 k}{(2\pi)^3} \Theta \left(\tilde{\mu}_q - E_q (k)\right) \left[\tilde{\mu}_q - E_q (k) \right]- \sum_{q=[u,d]}\frac{V_0^2}{8G_v},  
\end{eqnarray} 
where the first term has divergence, therefore it has to be regularized using the  { PTR scheme}. The reduced quark chemical  {potentials for $u$ and $d$ quarks are} defined by $\tilde{\mu}_u = \tilde{\mu}_d = \mu_u - 2 G_v \rho^v_u$, as also given in Eq.~(\ref{qnjl5}). $V_0 = 2 G_v \langle \psi_q^\dagger \psi_q \rangle$, $E_u (k) = E_d (k) = \sqrt{\vec{k}^2 + M^2_0 }$, and $M_{0}$ are the vector field (potential), quark energy, and the constituent quark mass in free space, respectively. We note that we have subtracted a constant (free space) contribution ($M_q=M_0$) in the effective potential in Eq.~(\ref{qnjl6}). Therefore pressure  {becomes} zero in free space. Along with this effective potential definition and the so-called Gibbs-Duhem relation, the energy density will also vanish in free space~\cite{Ripka:1997zb}.

Using the effective potential in Eq.~(\ref{qnjl6}) the pressure and the energy density for the normal (nonstrange)  {QM} are respectively expressed by
\begin{eqnarray}
    \label{qnjl7}
        P_{\rm QM} &=& - \mathscr{V}^{\textrm{QM}}_{\textrm{NJL}} (M_q,\mu_q)-\mathscr{V}_l (\mu_l), \,\,\,\,\,\,
        E_{\rm QM} = \mathscr{V}^{\textrm{QM}}_{\textrm{NJL}} (M_q,\mu_q) + \sum_{q=[u,d]} \mu_q \rho^v_q,  \\
        \rho_q &=& - \frac{\partial \mathscr{V}^{\textrm{QM}}_{\textrm{NJL}} \left( M_q, \mu_q\right)}{\partial \mu_q}. 
\end{eqnarray}

\begin{figure}
  \begin{center}
    \includegraphics[width=0.95\textwidth]{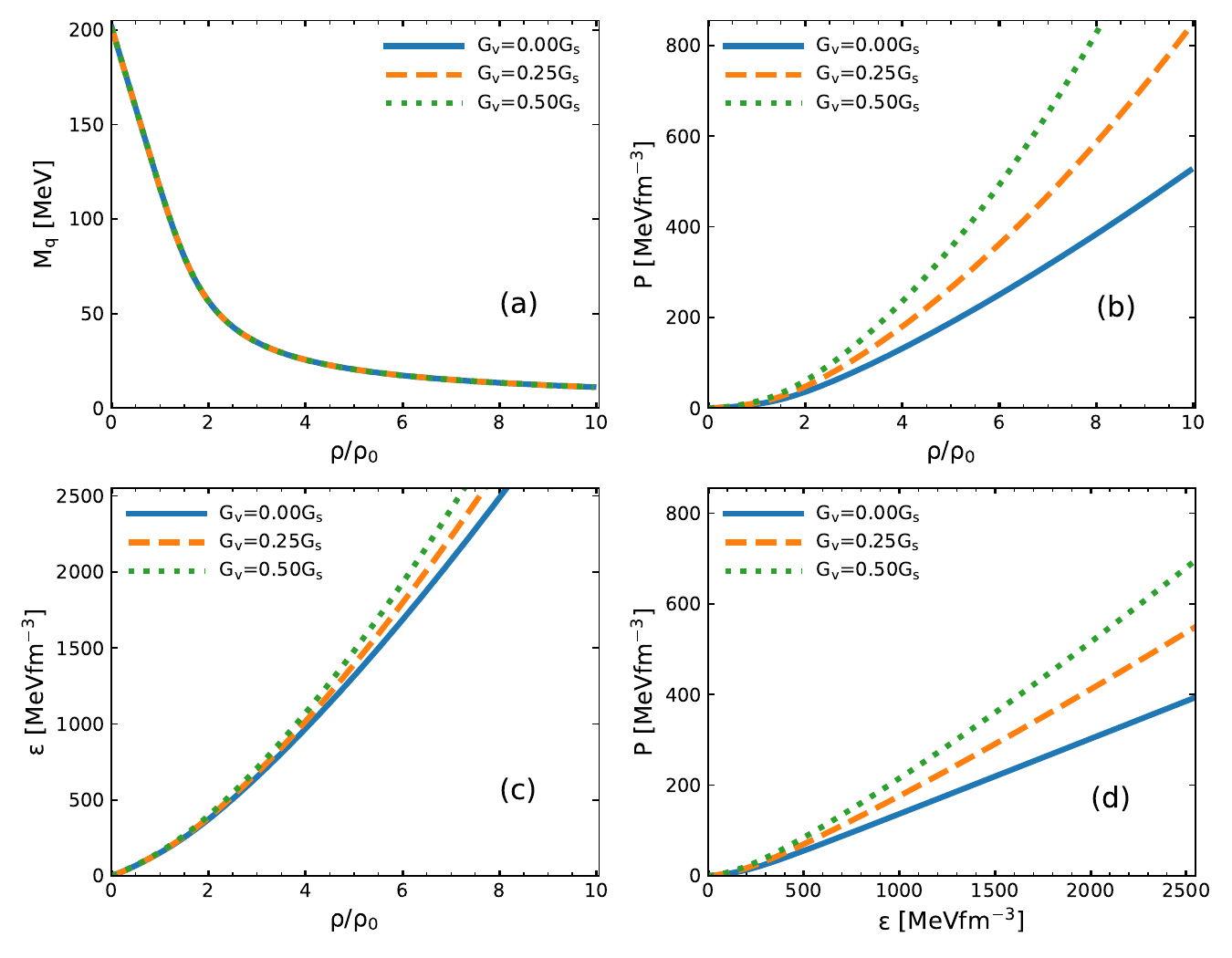}
  \end{center}
  \caption{From the NJLPTR model: (a) Effective quark mass for different values of $G_v$ as a function of $\rho/\rho_0$, (b) Pressure of QM as a function of $\rho/\rho_0$, (c) Energy density of QM as a function of $\rho/\rho_0$, and (d) EoS of PQM.}
  \label{fig2}
\end{figure}

In the calculation of the QM EoS, the NJL parameters have been fitted to the pion mass $m_\pi =  0.14$ GeV, and the pion weak decay constant $f_\pi =$ 0.093 GeV, giving  $G_{s} =$ 3.17 GeV$^{-2}$ and $\Lambda_{\mathrm{UV}} =$ 1.0789 GeV. The constituent quark mass  {$M_0$} = 0.20 GeV and the current quark mass $m_q =$ 0.0055 GeV, which is similar to PDG values~\cite{ParticleDataGroup:2022pth}. Results for the effective quark mass, pressure, and energy density of the quark matter for vector interaction couplings $G_v = (0.00, 0.25, 0.50) G_s$ are shown in Fig.~\ref{fig2}. As expected, the effective quark masses decrease as the density increases as shown in Fig.~\ref{fig2} (a). Additionally, the effective quark mass is not sensitive to the change of the values of the vector repulsive interaction coupling, where the rate of change is slightly low. This effective quark mass result is consistent with the result obtained in Refs.~\cite{Baym:2017whm,Lawley:2005ru}.

Results for the pressure of the PQM are given in Fig.~\ref{fig2} (b). It shows that the pressure is stiffer with a larger value of $G_v$.
Such  {behavior} is a natural and expected result because a larger $G_v$ value makes the repulsive force  {become} stronger. In addition, we show the results for the $E$-$\rho$ relation of the PQM for different values of $G_v$ in Fig.~\ref{fig2} (c). 
A similar result with the pressure is found:  {The} energy density increases as the $G_v$ value and density increase, which can be clearly understood from the relation of $P_{\rm QM} = - \mathscr{V}^{\rm QM}_{\rm NJL} - \mathscr{V}_l$ and  {$E_{\rm QM} = \mathscr{V}^{\rm QM}_{\rm NJL} + \sum_q \mu_q \rho_q^\nu$}, as given in Eq.~(\ref{qnjl7}). It implies that increasing the value of pressure is only given by increasing the energy density (increasing $-E$) and \textit{vice versa}, as explained in the visualization of Fig.~12 of Ref.~\cite{Baym:2017whm}. Figure~\ref{fig2} (d) shows EoS $P$-$E$ relation for different values of $G_v$. We found that the pressure increases as the energy density and $G_v$ increase. This also indicates that the EoS of the QM is stiffer for higher values of $G_v$.

\subsection{Properties of the static hybrid star} \label{sec:PNS}
\begin{figure}
  \begin{center}
    \includegraphics[width=0.95\textwidth]{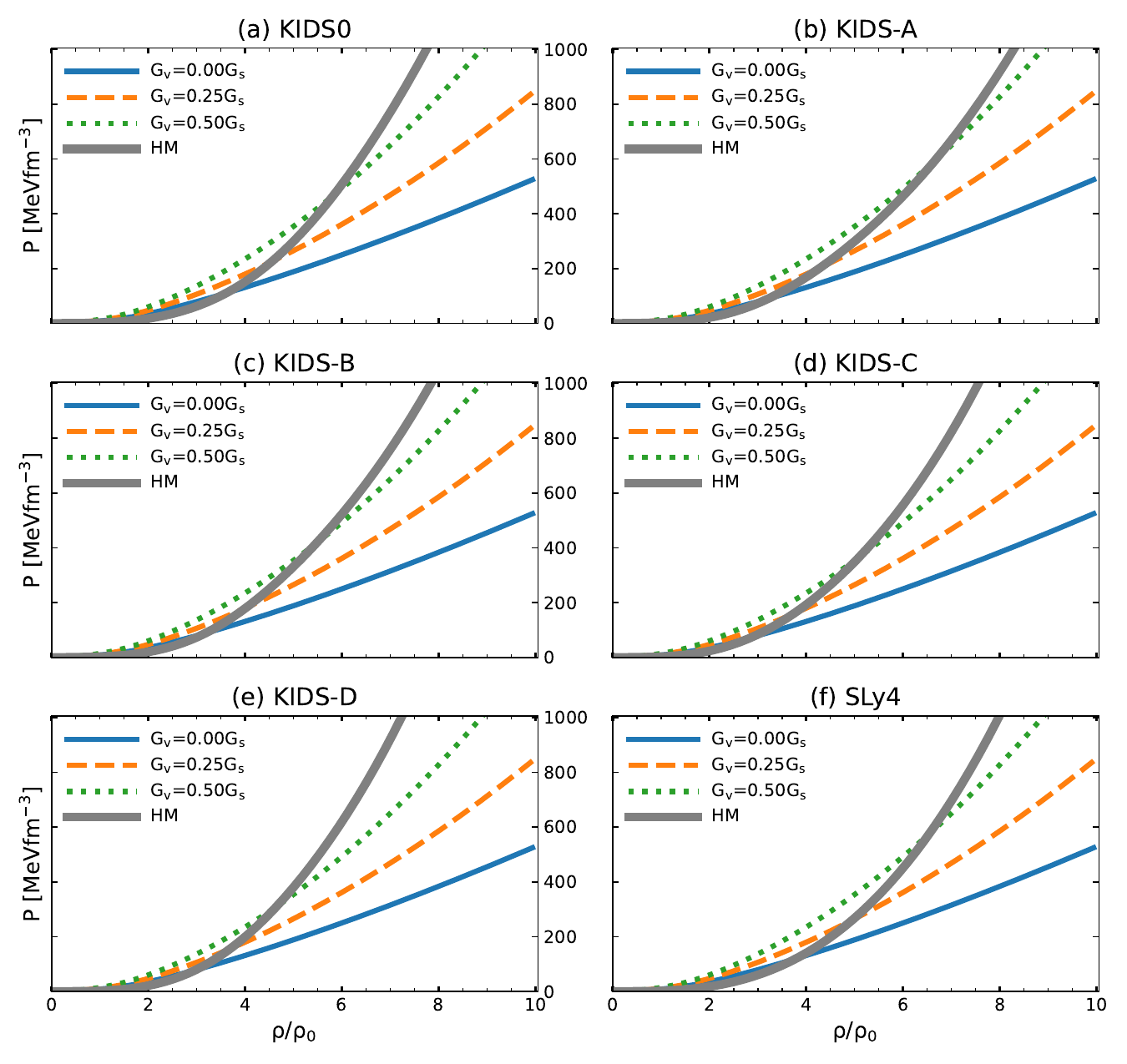}
  \end{center}
  \caption{Pressure as a function of density for QM with different values of $G_v$ and for HM in the different EDF models: (a) KIDS0, (b) KIDS-A, (c) KIDS-B, (d) KIDS-C, (e) KIDS-D, and (f) SLy4. The critical density $\rho_c$ is given by the cross point between $P_{\rm HM}$ and $P_{\rm QM}$ with different values of $G_v$. The thick solid line is $P$-$\rho$ relation for HM. The thin solid line is $P$-$\rho$ relation for QM with $G_v = 0.00 G_s$, the dashed line is $P$-$\rho$ relation for QM with $G_v = 0.25 G_s$, and the dashed line is $P$-$\rho$ relation for QM with $G_v = 0.50 G_s$.}
  \label{fig3}
\end{figure}

Here, we input the  {obtained} EoSs for hadronic and quark matters for different values of $G_v$ and symmetry energies into the TOV equation to numerically compute the properties of a non-rotating NS, obtaining the mass-radius (M-R) relation (structure of the hybrid star).  The TOV equations are given by~\cite{Oppenheimer:1939ne,Tolman:1939jz,Tolman:1934za}
\begin{eqnarray}
    \frac{dP(r)}{dr} &=& - \frac{G[E(r) + P(r)] [M(r) + 4\pi r^3 P(r)]}{r[r-2GM(r)]}, \,\,\, \, \, \, \, \frac{dM(r)}{dr} = 4\pi r^2 E(r),
\end{eqnarray}
where $P(r)$ and  {$E(r)$} are respectively the pressure and energy density at radial position $r$. $G$ and $M(r)$ are the gravitational constant and mass within the sphere of radius $r$, respectively. Using the EoSs of the constructed hybrid stars as input, we obtain the $M$-$R$ relation as a result.  {Note that the EoSs of the hybrid stars are depicted in Fig.~\ref{fig3}}.

\section{Numerical result} \label{sec:NR}

\subsection{Critical density of  {the hybrid star}}
In this section, we present the results  {of the critical densities of the phase transition of the hadronic to the quark matter for different KIDS models and the Skyrme model}. The critical density $\rho_c$ is simply determined by finding the cross point between $P_{\rm HM}$ and $P_{\rm QM}$.  {It is also worth noting that the $\rho_c$ is the phase transition from the hadronic and quark phases that occurs when  $P_{\rm HM}(\rho_c) = P_{\rm QM}(\rho_c)$ and $\rho_{\rm HM} = \rho_{\rm QM}$}. In Fig.~\ref{fig3}, it shows $P_{\rm HM}$ for each hadronic model and $P_{\rm QM}$ with three values of the vector coupling constant $G_v = 0$, $0.25G_s$, and $0.5G_s$. Figure~\ref{fig3} (a) shows $P_{\rm HM}$ for the KIDS0 model and $P_{\rm QM}$ for the NJL model with different values of $G_v$. The critical density for $P_{\rm HM}$ and $P_{\rm QM}$ for $G_v =0.00 G_s$ is around $\rho_c =3.60 \rho_0$. The $\rho_c$ increases as the value of $G_v$ increases. The $\rho_c$ for the KIDS-A model is shown in Fig.~\ref{fig3} (b). Note that the KIDS-A model has stiffer symmetry energy than that for the KIDS0 model. The critical density for the KIDS-A model and the NJL model with $G_v = 0.00 G_s$ and $G_v=0.25 G_s$ are smaller than those for the KIDS0 model. However, the critical density for the KIDS-A model and the NJL model with $G_v = 0.5 G_s$ is larger than that obtained for the KIDS0 model. This feature is followed by the KIDS-B, KIDS-C, and KIDS-D models as shown in Figs.~\ref{fig3} (c)-(e), while the SLy4 model and the NJL model with different values of $G_v$ almost have similar critical density values as those for the KIDS0 model and the NJL model with the corresponding values of $G_v$. This can be understood because both KIDS0 and SLy4 models have softer nuclear symmetry energies in comparison to other KIDS models as shown in Fig.~\ref{fig1}. For all the hadronic models and the NJL model with different $G_v$ values, the EoS is stable in the hadron phase at low densities, and the QM becomes more stable than the HM at high densities.

Again, comparing the $\rho_c$ values for larger values of $G_v=0.5G_s$ among the KIDS-A, KIDS-B, KIDS-C, and KIDS-D models,
 {the highest value of $\rho_c$ is given by the KIDS-A model}, and it decreases in the order of KIDS-B,  {KIDS-C, and KIDS-D models}. 
The ordering of $\rho_c$ could be understood from the stiffness of symmetry energy. With a soft symmetry energy (small $L$ and $K_{\rm sym}$ values, where $L$ and $K_{\rm sym}$ are respectively the slope and the curvature of symmetry energies),
the energy to create a neutron becomes small, so the $\beta$-equilibrium condition allows easy creation of the neutron, which eventually leads to a large fraction of the neutron. Pauli blocking makes the pressure exerted by the neutron strong when there are more neutrons within the neutron star. For this reason, the pressure becomes stiff from KIDS-A to KIDS-D while the symmetry energy is stiffened in the reverse order. Exact values of $\rho_c$ are summarized in Tab.~\ref{tab1}. One can see that, regardless of $G_v$ value, the transition to the quark phase occurs at low densities with soft symmetry energy. Values of the pressure at $\rho_c$ ($P_c$) are also summarized in the table.

Overall, it indicates that the phase transition is highly sensitive to both the nuclear symmetry energy in the hadron phase and the vector repulsion in the quark phase. Their effect on physical observables can be probed by solving the TOV equations and obtaining the mass and radius of neutron stars.

\begin{table}[tbp]
    \centering
       \caption{ \label{tab1} Transition point obtained from the $P$-$\rho/\rho_0$ relation for the hybrid KIDS-NJLPTR models with different values of $G_v$. The units of the coupling constants of $G_v$ and $G_s$ are MeV$^{-2}$, and $P$ is MeV$\cdot$fm$^{-3}$. Note that $\rho_c$ is in the unit of $\rho_0$.}
    \begin{tabular}{c|c|c|c|c|c|c}  \hline
           $G_v$ & KIDS0 & KIDS-A & KIDS-B & KIDS-C & KIDS-D & SLy4  \\ \hline \hline
            & ($P_c,\rho_c$) & ($P_c,\rho_c$) & ($P_c,\rho_c$) & ($P_c,\rho_c$) & ($P_c,\rho_c$) & ($P_c,\rho_c$) \\ \hline
    $0.00 G_s$ & ($110.23,3.60$)  & ($89.98, 3.21$) & ($88.21, 3.18$) & ($73.00,2.86$) & ($78.15, 2.97$) & ($119.73, 3.78$) \\ 
    $0.25 G_s$ & ($226.44, 4.55$) & ($203.39, 4.29$) & ($181.44, 4.02$) & ($159.28, 3.74$)& ($155.71, 3.69$) & ($261.06, 4.95$) \\
    $0.50 G_s$ & ($464.75, 5.81$) & ($592.96, 6.65$) & ($417.06, 5.47$) & ($365.00, 5.09$) & ($309.08, 4.65$) & ($574.79, 6.54$) 
          \\ \hline
    \end{tabular}
\end{table}

\subsection{ {$M$-$\rho$ and $M$-$R$ relations of the hybrid star}}
\begin{figure}
  \begin{center}
    \includegraphics[width=0.95\textwidth]{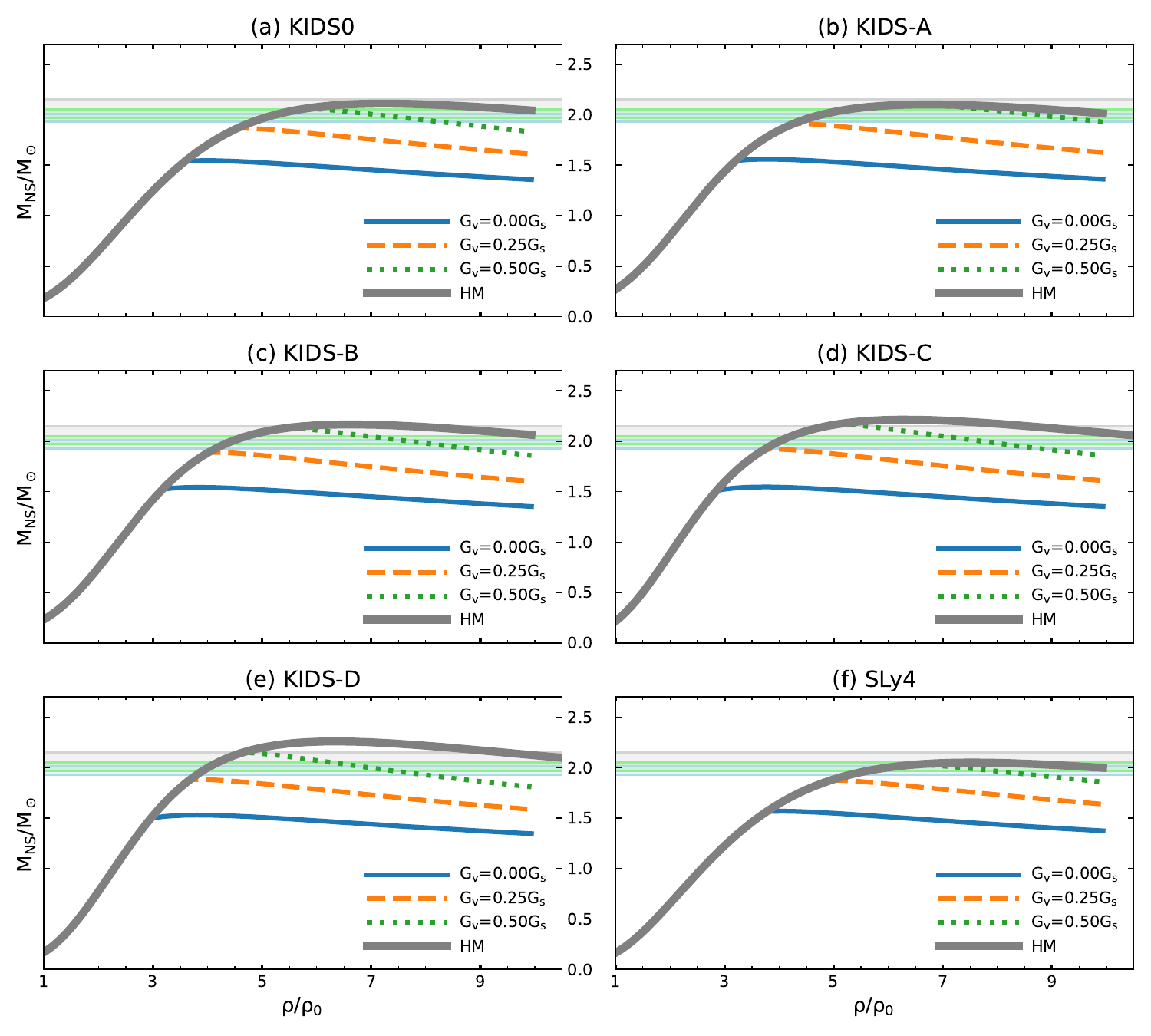}
  \end{center}
  \caption{NS mass-density relations of different hadron-quark matter EoS with different values of $G_v$ for different EDF models: (a) KIDS0, (b) KIDS-A, (c) KIDS-B, (d) KIDS-C, (e) KIDS-D, and (f) SLy4. The thick solid line represents the NS mass for the HM. The thin solid line is the NS mass for the HM+QM with $G_v = 0.00 G_s$, the dashed line is the NS mass for the HM+QM with $G_v = 0.25 G_s$, and the dotted line represents the NS mass for the HM+QM with $G_v = 0.50 G_s$.}
  \label{fig4}
\end{figure}

Figure~\ref{fig4} displays the mass of an {NS} as a function of the density at the center. For the pure  {hadronic} matter, all six models produce  {the NS} maximum masses ($M_{\rm max}$) greater than $2M_\odot$. The  {NS} maximum mass for the KIDS-D model is the largest as shown in Fig.~\ref{fig4} (e), and it decreases in the order of KIDS-C,  {KIDS-B, and KIDS-A models} as in Figs.~\ref{fig4} (b)-(d). This result is consistent with the stiffness of EoS in Fig.~\ref{fig1}. The result changes dramatically if  {QM} exists in the core of the star.

The thick solid lines denote the NS mass results of the hadronic matter as in Fig.~\ref{fig4} (a)-(f). In all the hybrid KIDS-NJL models, the density at which the thin solid lines begin is equal to $\rho_c$ in Tab.~\ref{tab1}, which means that as the transition to quark matter occurs, the EoS becomes soft then the matter cannot support strong gravity. As a result, $M_{\rm max}$ is determined at the density of phase transition. For $G_v=0$, $\rho_c$ is not much sensitive to the  {models}, so the NS maximum masses are obtained in a narrow range (1.5--1.6)$M_\odot$. These values are substantially low in comparison to $2M_\odot$, so the result confirms that vector repulsion must be necessarily accounted for in the QM to reproduce the observation of the large mass of NS. The dashed lines correspond to the result of $G_v=0.25G_s$ as shown in Fig.~\ref{fig4} (a)-(f). Similar to $G_v=0$, the density at which a dashed line begins is the same with $\rho_c$ in Tab.~\ref{tab1}, and $M_{\rm max}$ is determined at the density where the mass is below $2M_\odot$ for all EDF models. The result demands that the $G_v$ must be higher than $0.25 G_s$.

The results of $G_v=0.5G_s$ which are shown with dotted lines are now consistent with the $2M_\odot$ constraint. The density at which the QM curve begins is the highest is given by the KIDS-A model, and it then decreases in the order of KIDS-B,  {KIDS-C, and KIDS-D models}. This ordering is understood easily in terms of the stiffness of the EoS of each model. Contrary to the results of $G_v=0$ and $0.25G_s$, the density at which the dotted curves begin is higher than the $\rho_c$ values in Tab.~\ref{tab1}. The difference means that even after the QM is formed in the core, its EoS is stiff enough that it can resist gravitational contraction up to a certain density,
at which the NS star reaches the maximum value. However, the interval between $\rho_c$ and the density at the maximum mass of NS is narrow, so the fraction of QM in the core of the NS is not likely to be significant. Summarizing the result, to satisfy the $2M_\odot$ condition, the $G_v$ value must be larger than a certain value to obtain the QM EoS stiff. The stiff QM EoS increases $\rho_c$, and the high $\rho_c$ value constrains the existence of QM in a limited range.

\begin{figure}
  \begin{center}
    \includegraphics[width=0.95\textwidth]{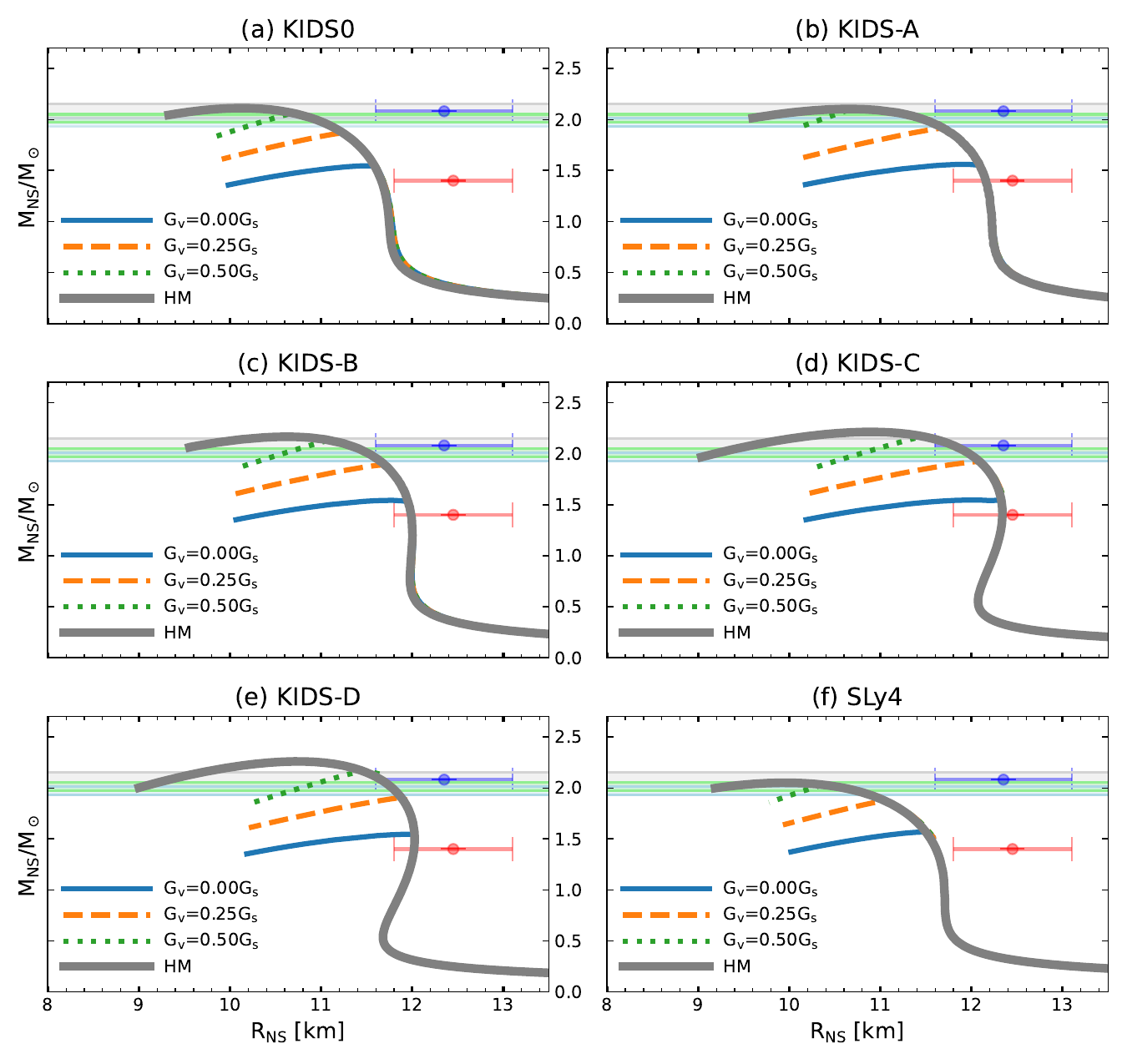}
  \end{center}
  \caption{Same as Fig.~\ref{fig4}, but for NS $M$-$R$ relations. Horizontal error bars show the radius of $1.4M_\odot$ and $2.08M_\odot$ stars determined in the NICER analysis~\cite{apjl918}.}
  \label{fig5}
\end{figure}

Figure \ref{fig5} presents the mass-radius ($M$-$R$) relation obtained from the TOV equations. Since the data from GW and NICER provide information on both mass and radius of NS, they can constrain the EoS of HM and QM more strictly than the data of mass  {alone}. The radius of NS canonical mass ($1.4M_\odot$) $R_{1.4}$ is of particular interest and importance. According to the NICER analysis~\cite{apjl918}, the NS radius is $R_{1.4} = 12.45 \pm 0.65$~km  within $1\sigma$ credible interval~\cite{apjl918}.
On the theoretical side, the density at the center of $1.4M_\odot$ NS is obtained not to exceed $3 \rho_0$ in general. The density at which exotic states such as the mixture of the hyperon or transition to QM appear is above $2\rho_0$ as a whole, though the uncertainty is non-negligible. Therefore, the properties of $1.4M_\odot$  {NS} are especially important to probe (i) the EoS of HM at high densities if there is no transition to QM or hyperon, or  (ii) consistency of the EoS of exotic phases with the data of $1.4M_\odot$  NS  if these exotic states exist in the interior of the $1.4M_\odot$ NS. The $M$-$R$ result shows that the effect of QM appears at  NS masses higher than $1.5M_\odot$, so the radius of $1.4M_\odot$ NS is unaffected by the transition to QM. Hyperon is another source that can soften the EoS of $1.4M_\odot$ NS. According to work in progress \cite{kidsll}, if the $\Lambda\Lambda$ interaction determined from the experimental data of double-$\Lambda$ hypernuclei is included in the EoS of hyperon matter, it makes the EoS stiff. As a consequence, the maximum mass of the NS becomes larger than $2M_\odot$, and the creation of the hyperon gives negligible effect to the radius of $1.4M_\odot$ star. Therefore, accurate measurement of $R_{1.4}$ can provide  {stringent} constraints on the EoS of HM at densities up to $3\rho_0$.

In comparison with other results, our results show the difference to some extent, but qualitatively both results are compatible. In works of Refs.~\cite{prd100, apj934},  they reported that transition or crossover to QM is considered. In both works, QM was described in terms of the NJL model, but for the HM, an RMF model is used in \cite{prd100}, and a chiral effective field theory ($\chi$EFT) and Togashi EoSs are employed in \cite{apj934}.

In Ref.~\cite{prd100}, the result of the MS-B+vNJL model is comparable to our work. Three values are assumed for the vector repulsion of QM, $G_v/G_s = 1.5$, 2.0, and 2.5. The result shows that the NS maximum mass is below $2M_\odot$ when $G_v=1.5G_s$, and it becomes obviously above $2M_\odot$ when $G_v= 2.0 G_s$. It is notable that, in our calculation, $M_{\rm max} \geq 2M_\odot$ is obtained if $G_v \geq 0.5 G_s$, so there is a huge difference in the strength of the vector repulsion.
 {$M$-$R$} curve of MS-B+vNJL is also very different from what we obtain. Transition to QM occurs at $\rho = 1.5\rho_0$ or $\rho=2.0\rho_0$  in the MS-B+vNJL model, and then the radius becomes smaller than the radius of the HM.
The radius of the $1.4M_\odot$ NS is about 12.5~km without QM, but it falls in the range 11.7--12.2~km depending on the $G_v$ and $\rho_c$ values. Radii of the maximum mass NSs are located below 11~km,  {which} is much smaller than the NICER estimation $R_{2.08} = 12.35 \pm 0.75$~km, where $R_{2.08}$ represents the radius of $2.08 M_\odot$ NS \cite{apjl918}.

In Ref.~\cite{apj934}, the crossover between  HM to QM is assumed to begin at $1.5\rho_0$, and the dependence
on $G_v$ is presented. Compared to the $R_{1.4}$ of Togashi EoS in which there are only nucleons, the  {NS} radius becomes larger if the  {QM} is included in the core by about 0.2--0.3~km. This behavior is  {opposite} to the result of MS-B+vNJL in \cite{prd100}.
The radius of $2.08M_\odot$  {NS} is in the range 11.5--12~km for the $G_v$ values that have maximum masses larger than $2.08M_\odot$. It is  {shown} in the paper that  {the values of $G_v$ must be} larger than $0.84 G_s$ to satisfy the condition 
$M_{\rm max} \geq 2.08 M_\odot$. Also, this shows that the $G_v$ value in  {Ref.~\cite{apj934} is much larger than the value of $G_v$ that is obtained in our work to satisfy the $2M_\odot$}. However, some recent references in the  iterature~\cite{Alaverdyan:2022foz,Kumar:2023lhv} also reported that the strength of $G_v$ is in the same order as that used in this work to give $2M_\odot$. It is worth noting that there is no stringent constraint for the values of $G_v$ available nowadays. More precise data are needed to constrain $G_v$.

It is informative to compare our result of  {NS} radius around the NS maximum mass with the NICER result in $R_{2.08}=11.6$--$13.1$~km. With  {HM} only, the  {NS} radii at NS maximum mass are in the range of 9.8--11~km.  {The NS} radii tend to increase in a stiffer EoS,  {giving} the largest  {NS} radius for the KIDS-C model. However, even the largest value of  {NS radius} is  {relatively} smaller  {compared to} the lower limit of NICER $R_{2.08}$. The inconsistency could be understood as a signal for the existence of phases other than the  {hadrons}. Comparing the results of KIDS-A,  {KIDS-B, KIDS-C, and KIDS-D models},  a soft EoS of  {HM} gives larger values of $\rho_c$, so even if the  {QM} is formed in the core, $R_{\rm max}$ (radius of the NS maximum mass)
is too small to be consistent with the range of NICER $R_{2.08}$. The KIDS-A model clearly shows such behavior. On the other hand, stiff EoSs like  {the} KIDS-C and D  {models} show that the phase transition occurs at a relatively large NS radius, so the  {NS} radius of the maximum mass is  {shifted} to the values larger than those of pure HM. As a result, the NS with QM in the core can explain the NICER $R_{2.08}$ data better than the  {NSs} with nucleons only.

While the KIDS-A,  KIDS-B, KIDS-C, and KIDS-D models are determined to satisfy the constraint $R_{1.4}=$ 11.8-12.5~km, which is more restrictive than the NICER $R_{1.4}$,  {the} SLy4 and KIDS0 models are fitted to the APR pure neutron matter EoS \cite{apr} in the isovector component of the functional. For this reason, the  {$M$-$R$} behaviors of  {the} SLy4 and KIDS0 models are similar to each other, which are not consistent with NICER $R_{1.4}$ and $R_{2.08}$.

The result of  {this} work clearly shows that the symmetry energy plays an important role in not only controlling the stiffness of
EoS of HM, but also affecting the bulk properties of the NSs with QM core. The KIDS-C and D models with $G_v$ values larger than $0.5G_s$ satisfy the NS mass data of $1.4M_\odot$ and $2.08 M_\odot$. The values of the nuclear symmetry energy parameters $L$ and $K_{\rm sym}$ are respectively 58 and $-91.5$ MeV for the KIDS-C model, and they are 47 and $-134.5$ MeV for the KIDS-D model, respectively. These values are consistent with the 95~\% credible range obtained from the KIDS-R14 model set \cite{npsm2022}, $L=49.8 \pm 10.4$ MeV, and $K_{\rm sym} = -82.4 \pm 67.4$ MeV.

\section{Summary} \label{sec:SUM}
In summary, the role of nuclear symmetry energy has been investigated in the transition from hadron matter 
to deconfined quark phase in the NS core. Major concerns are (i) the dependence of the critical density for the phase transition on the symmetry energy and (ii) the role of the repulsive vector coupling in the EoS of QM and NS properties. The nuclear symmetry energy is determined to satisfy the data of the canonical mass of NSs. The effect of the uncertainty due to the symmetry energy becomes obvious in the $M$-$R$ relation of the  {NSs with} masses close to $2M_\odot$.

Quark matter EoS is described in the NJL model. For a rigorous treatment of the ultraviolet divergence, we employed the PTR scheme.
It is known that repulsive vector coupling is essential to reach the maximum mass of the hybrid star  {to be} consistent with the $2M_\odot$ astrophysical observation. Dependence on the vector coupling is examined by using three  {different} values of the vector coupling constants $G_v =$ 0.00, 0.25, and 0.50 in the unit of the $G_s$. Critical density for the phase transition is determined from the condition  $P_{\rm HM}(\rho_c) = P_{\rm QM}(\rho_c)$ at $\rho_{\rm HM} = \rho_{\rm QM}= \rho_c$.

We found that the critical density is highly sensitive to symmetry energy. Comparing the $\rho_c$ values among the four models: KIDS-A,  KIDS-B, KIDS-C, and KIDS-D models, the value  {of $\rho_c$} tends to decrease with softer symmetry energy, i.e., $\rho_c({\rm A}) > \rho_c({\rm B}) > \rho_c({\rm C}) > \rho_c({\rm D})$ regardless of the vector coupling constant. When the vector repulsion is turned off ($G_v=0$), the maximum mass  {NS} is  {obtained about} (1.5--1.6)$M_\odot$. This confirms that repulsive vector coupling is  {really needed} for obtaining the consistency result with $2M_\odot$ observation. Since the NS maximum masses are obtained at $\rho_c$ for $G_v=0$, properties of the $1.4M_\odot$ NSs are not affected by the transition to QM. Therefore, the EoS of $1.4M_\odot$ NS could be determined accurately in terms of the hadronic degrees of freedom.

For $G_v=0.25G_s$,  {once the QM} is created in the NS core, the NS reaches the maximum mass around $1.9M_\odot$, which is quite independent of the model. To satisfy the observation, a stronger repulsion is required. With $G_v=0.50G_s$, we found that the {NS} maximum mass is consistent with astrophysical observation, but the  {NS} radius becomes inconsistent with the observation findings. For the KIDS-A and B models, the phase transition does not improve the  {NS $M$-$R$} relation, which is out of the range of the NS radius determined by the NICER analysis. On the other hand, the KIDS-C, and D models agree with the NICER range if the QM exists in the NS core.

Consequently, we have shown that the density dependence of the symmetry energy and $G_v$ play a critical role in the phase transition from hadron to quark matters. Critical density is sensitive to both symmetry energy and the vector coupling constant. Their effects are crucial to the  {$M$-$R$} behavior of the hybrid stars whose masses are above the NS canonical mass. Precise measurement of the large mass of the NSs will offer a unique opportunity to constrain the symmetry energy in the HM and repulsion vector coupling in the  QM simultaneously.

\section*{Acknowledgments}
P.T.P.H. thanks to Daniel Whittenbury for the discussion. This work was supported by the National Research Foundation of Korea (NRF) Grant Nos.~2018R1A5A1025563, 2022R1A2C1003964, 2022K2A9A1A0609176, and 2023R1A2C1003177.



\begin{thebibliography}{99}
\bibitem{epja2022}
S.~Choi, E.~Hiyama, C.~H.~Hyun and M.~K.~Cheoun, 
Eur. Phys. J. A \textbf{58}, no.8, 161 (2022).

\bibitem{kcond1}
C.~Y.~Ryu, C.~H.~Hyun, S.~W.~Hong and B.~T.~Kim, 
Phys. Rev. C \textbf{75}, 055804 (2007).

\bibitem{kcond2}
Y.~Lim, K.~Kwak, C.~H.~Hyun and C.~H.~Lee, 
Phys. Rev. C \textbf{89}, no.5, 055804 (2014).

\bibitem{Papakonstantinou:2016zpe}
P.~Papakonstantinou, T.-S.~Park, Y.~Lim and C.~H.~Hyun, 
Phys. Rev. C \textbf{97}, no.1, 014312 (2018).

\bibitem{kids-nuclei1}
H.~Gil, P.~Papakonstantinou, C.~H.~Hyun and Y.~Oh, 
Phys. Rev. C \textbf{99}, no.6, 064319 (2019).

\bibitem{kids-nuclei2}
H.~Gil, Y.-M.~Kim, C.~H.~Hyun, P.~Papakonstantinou and Y.~Oh, 
Phys. Rev. C \textbf{100}, no.1, 014312 (2019).

\bibitem{kids-sym}
H.~Gil, Y.-M.~Kim, P.~Papakonstantinou and C.~H.~Hyun, 
Phys. Rev. C \textbf{103}, no.3, 034330 (2021).

\bibitem{kids-k0}
H.~Gil and C.~H.~Hyun, 
New Phys.: Sae Mulli \textbf{71}, no.3, 242-248 (2021).

\bibitem{kids-ksym}
H.~Gil, P.~Papakonstantinou and C.~H.~Hyun, 
Int. J. Mod. Phys. E \textbf{31}, no.01, 2250013 (2022).

\bibitem{Hutauruk:2022bii}
P.~T.~P.~Hutauruk, H.~Gil, S.~i.~Nam and C.~H.~Hyun, 
Phys. Rev. C \textbf{106}, no.3, 035802 (2022).

\bibitem{Hutauruk:2022bso}
P.~T.~P.~Hutauruk, H.~Gil, S.~i.~Nam and C.~H.~Hyun,
PTEP \textbf{2023}, no.6, 063D01 (2023).

\bibitem{Danielewicz:2002pu}
P.~Danielewicz, R.~Lacey and W.~G.~Lynch, 
Science \textbf{298}, 1592-1596 (2002).

\bibitem{Tews:2012fj}
I.~Tews, T.~Kr\"uger, K.~Hebeler and A.~Schwenk, 
Phys. Rev. Lett. \textbf{110}, no.3, 032504 (2013).

\bibitem{Antoniadis:2013pzd}
J.~Antoniadis, P.~C.~C.~Freire, N.~Wex, T.~M.~Tauris, R.~S.~Lynch, M.~H.~van Kerkwijk, M.~Kramer, C.~Bassa, V.~S.~Dhillon and T.~Driebe, \textit{et al.}
Science \textbf{340}, 6131 (2013).

\bibitem{NANOGrav:2019jur}
H.~T.~Cromartie \textit{et al.} [NANOGrav],
Nature Astron. \textbf{4}, no.1, 72-76 (2019).

\bibitem{Demorest:2010bx}
P.~Demorest, T.~Pennucci, S.~Ransom, M.~Roberts and J.~Hessels,
Nature \textbf{467}, 1081-1083 (2010).

\bibitem{Hutauruk:2021kej}
P.~T.~P.~Hutauruk and S.~i.~Nam, 
Phys. Rev. D \textbf{105}, no.3, 3 (2022).

\bibitem{Tanimoto:2019tsl}
T.~Tanimoto, W.~Bentz and I.~C.~Clo\"et, 
Rev. C \textbf{101}, no.5, 055204 (2020).

\bibitem{Hutauruk:2021dgv}
P.~T.~P.~Hutauruk and S.~i.~Nam, 
Mod. Phys. Lett. A \textbf{37}, no.14, 2250087 (2022).

\bibitem{Bentz:2001vc}
W.~Bentz and A.~W.~Thomas, 
Nucl. Phys. A \textbf{696}, 138-172 (2001).

\bibitem{Baym:2017whm}
G.~Baym, T.~Hatsuda, T.~Kojo, P.~D.~Powell, Y.~Song, and T.~Takatsuka, 
Rept. Prog. Phys. \textbf{81}, no.5, 056902 (2018).

\bibitem{Buballa:2003qv}
M.~Buballa, 
Phys. Rept. \textbf{407}, 205-376 (2005).

\bibitem{Glendenning:1992vb}
N.~K.~Glendenning, 
Phys. Rev. D \textbf{46}, 1274-1287 (1992).

\bibitem{Hutauruk:2022sbm}
P.~T.~P.~Hutauruk,
[arXiv:2204.11520 [hep-ph]].

\bibitem{Hutauruk:2016sug}
P.~T.~P.~Hutauruk, I.~C.~Cloet and A.~W.~Thomas,
Phys. Rev. C \textbf{94}, no.3, 035201 (2016).

\bibitem{Hutauruk:2018zfk}
P.~T.~P.~Hutauruk, W.~Bentz, I.~C.~Clo\"et and A.~W.~Thomas,
Phys. Rev. C \textbf{97}, no.5, 055210 (2018).

\bibitem{Schwinger:1951nm}
J.~S.~Schwinger,
Phys. Rev. \textbf{82}, 664-679 (1951)

\bibitem{Hell:2014xva}
T.~Hell and W.~Weise,
Phys. Rev. C \textbf{90}, no.4, 045801 (2014).

\bibitem{Klimt:1990ws}
S.~Klimt, M.~F.~M.~Lutz and W.~Weise,
Phys. Lett. B \textbf{249}, 386-390 (1990).

\bibitem{Lawley:2005ru}
S.~Lawley, W.~Bentz and A.~W.~Thomas,
Phys. Lett. B \textbf{632}, 495-500 (2006).

\bibitem{Ripka:1997zb}
G.~Ripka,
``Quarks bound by chiral fields: The quark-structure of the vacuum and of light mesons and baryons" (Oxford University Press, Oxford, 1997).

\bibitem{ParticleDataGroup:2022pth}
R.~L.~Workman \textit{et al.} [Particle Data Group],
PTEP \textbf{2022}, 083C01 (2022).

\bibitem{Oppenheimer:1939ne}
J.~R.~Oppenheimer and G.~M.~Volkoff,
Phys. Rev. \textbf{55}, 374-381 (1939).

\bibitem{Tolman:1939jz}
R.~C.~Tolman,
Phys. Rev. \textbf{55}, 364-373 (1939).

\bibitem{Tolman:1934za}
R.~C.~Tolman,
Proc. Nat. Acad. Sci. \textbf{20}, 169-176 (1934).

\bibitem{Macher:2004vw}
J.~Macher and J.~Schaffner-Bielich, 
Eur. J. Phys. \textbf{26}, 341-360 (2005).

\bibitem{Endo:2011em}
T.~Endo, 
Phys. Rev. C \textbf{83}, 068801 (2011).

\bibitem{apjl918}
M.~C.~Miller, {\it et al.}, 
Astrophys. J. Lett. {\bf 918}, L28 (2021).

\bibitem{kidsll}
S.~Choi, E.~Hiyama, C.~H.~Hyun, and M.-K.~Cheoun,
in progress.

\bibitem{prd100}
S.~Han, M.~A.~A.~Mamun, S.~Lalit, C.~Constantinou, and M.~Prakash,
Phys. Rev. D {\bf 100}, 103022 (2019).

\bibitem{apj934}
T.~Kojo, G.~Baym, and T.~Hatsuda,
Astrophys. J. {\bf 934}, 46 (2022).

\bibitem{Alaverdyan:2022foz}
G.~B.~Alaverdyan,
Astrophysics \textbf{65}, no.2, 278-295 (2022).

\bibitem{Kumar:2023lhv}
A.~Kumar, V.~B.~Thapa and M.~Sinha,
Phys. Rev. D \textbf{107}, no.6, 063024 (2023)

\bibitem{apr}
A.~Akmal, V.~R.~Pandharipande, and D.~G.~Ravenhall,
Phys. Rev. C {\bf 58}, 1804 (1998).

\bibitem{npsm2022}
C.~H.~Hyun,
New Phys.: Sae Mulli {\bf 72}, no.5, 371-375 (2022).
\end{thebibliography}
\end{document}